# Enforced freedom: electric-field-induced declustering of ionic-liquid ions in the double layer

*Yufan Zhang,* [†] *Ting Ye,* [⊥] *Ming Chen,* [⊥] *Zachary A. H. Goodwin,* [§,‡,∥] *Guang Feng,* [\*,⊥] *Jun Huang* [\*,╋], *and Alexei A. Kornyshev* [\*,§,∥]

† *IEK-13, Institute of Energy and Climate Research, Forschungszentrum Jülich GmbH, 52425 Jülich, Germany*

⊥ *State Key Laboratory of Coal Combustion, School of Energy and Power Engineering, Huazhong University of Science and Technology (HUST), Wuhan 430074, China*

§ *Department of Chemistry, Imperial College London, Molecular Sciences Research Hub, White City Campus, London W12 0BZ, U.K.*

‡ *Department of Physics, Imperial College London, South Kensington Campus, London SW7 2AZ, U.K.*

∥ *Thomas Young Centre for Theory and Simulation of Materials, Imperial College London, South Kensington Campus, London SW7 2AZ, U.K.*

╋ *Hunan Provincial Key Laboratory of Chemical Power Sources, College of Chemistry and Chemical Engineering, Central South University, Changsha 410083, China.*

\* E-mail: [gfeng@hust.edu.cn](gfeng@hust.edu.cn) (G.F.)

[jhuangelectrochem@qq.com](jhuangelectrochem@qq.com) (J.H.)

[a.kornyshev@imperial.ac.uk](a.kornyshev@imperial.ac.uk) (A.K.)

# Abstract

Whereas the majority of ions in the bulk of a solvent-free ionic liquid is bound into clusters, this is expected to change in the electrical double layer (EDL), in which the resulting electric field 'prefers' to interact with electrical monopoles – free, unclustered ions. The competition between the propensity of ions to stay in a clustered state and the reduction of the energy of ions in electric field in the free state determines the resulting portion of free ions in the EDL. We present a study of this effect, based on the simplest possible mean-field theory. 'Cracking' of ion clusters into individual ions in electric field is accompanied by the change of the dielectric response of ionic liquid which is different in clustered and unclustered states. The predictions of the theory are verified and further explored by specially performed molecular dynamics simulations. A particular finding of the theory is that the differential capacitance vs potential curve displays a bell shape despite low concentration of free charge carriers, because the dielectric response of bound ions reduces the threshold concentration of the bell-to camel-shape transition. Whereas qualitatively these findings make perfect sense, in reality the exact numbers and criteria might be different as the presented simple theory does not take into account overscreening and oscillating charge and electrostatic potential distributions near the electrode. This is why testing the theory with computer simulations is essential, but the latter basically reproduce the qualitative conclusions of the theory.

## Introduction

Electrochemical double layer capacitors, or supercapacitors, store energy by forming oppositely charged electrical double layers (EDLs) at the cathodes and anodes.[1] They have significantly higher power density than conventional Li-ion batteries, because their charging process does not involve any electrochemical reactions, but just a rearrangement of cations and anions.[2] Owing to that they also have significantly longer life time, sustaining millions of charging-discharging cycles, versus cycle life on the magnitude of $10^3$ for Li-ion batteries.[3] One of the priorities in supercapacitor research is the increase of energy density which is inferior to those of batteries. One way to rectify this is by massively increasing of the surface area of the electrode | electrolyte interfaces, by using micro- and mesoporous electrodes. Another avenue for energy increase is the use of electrolytes that could sustain higher applied voltages without undergoing electrochemical reactions. The use of ionic liquids (ILs), with their excellent thermal stability, nonvolatility, and relatively inert nature, enables one to attain higher operation voltages than aqueous and even organic electrolytic solutions and thereby helps to store more energy.[4-6] Accurate theoretical description of the behavior of ILs in the EDL and, specifically of charge distribution and double layer capacitance, is a premise to understand their performance in supercapacitors (for review see Refs. 7 and 8).

Many modified Poisson-Boltzmann (PB) theories of the EDL, of different levels of complexity, have been put forward.[9-21] The majority of them modeled the system as ions embedded in a continuum medium with constant permittivity, which, as first pointed out by Debye, is insufficient to depict reality because the dielectric saturation effect on solvent polarizability is not considered.[22] Onsager, Kirkwood and

Booth developed theories to account for this issue, that involved the reduction of the dielectric constant of the solvent in the vicinity of electrode.[23-26] The dielectric constant ($\epsilon_d$) affects how efficiently ions could screen the electric field, and therefore it is closely related to the properties of the EDL, such as the differential capacitance ($C$).[26-29] The reduced dielectric response of the solvent would generally result in a decreased $C$ at the electrode-electrolyte interface. Hence, more recently, researchers have modified their models to include the permanent dipoles and the excluded volume of solvents and ions, as well as the solvent quadrupolarizability.[30-35]

Little attention, however, has been put into the variation of the dielectric response in neat ILs with transient ion-clustering.[36, 37] In such solvent-free systems, the dielectric constant is predominantly contributed by transient ionic clusters, and by the dipole moments of ions themselves. Whereas Ref. 36 has studied the balance of and interchange between free and clustered ionic states, it is equally important to study equilibria between such states in EDL, which is the subject of the present article.

In this work, we formulate a mean-field lattice-gas model of the EDL in a simplified IL system with consideration of the dipole moment of ion pairs, considering for simplicity the clustering at the ion pair level. This is, of course, a strong idealization, but it is a first step in this direction, which helps us to elucidate the main qualitative effect: 'liberation' (declustering) of ions in the strong electric field of the double layer. This effect and the whole idea of the field-induced declustering is in spirit of the old classical Damaskin-Frumkin-Parsons cluster model of the compact layer capacitance in aqueous electrolyte. But we will explore its manifestations in the whole double layer of ionic liquid electrolytes.[38, 39]

Firstly, we derive an analytical expression for the spatially-varying dielectric constant ($\epsilon_d$). Then, we describe the behavior of clustered ions in and out of EDL, followed by the comparison of the predicted number density distribution of clustered ions with the corresponding values obtained in molecular dynamics (MD) simulations. In addition, we investigate the behavior of differential capacitance as a function of electrode potential, demonstrating a new trend in the "camel-bell transition",[14] which is different from the prediction of the theory that does not take into account declustering in electric field.

The theory itself, should, of course, be taken with a 'pinch of salt': at best it can claim only qualitative results, as it does not incorporate the effect of overscreening and the decaying oscillations of charge density and electrical potential in the EDL.[7, 18] Therefore, the presented tests of its conclusions by MD-simulations become critical.

## Methods

*The model and mean-field theory*

We consider a system where ions exist in two states: free state and bound state, with interchange between them. For simplicity of analysis in this work, we assume that cations and anions in free state are of the same size, monovalent with charge $\pm e$ and do not possess permanent dipole moments (which, strictly speaking can be true only for a limited number of ILs),[7] with their electronic polarizability determining an effective dielectric constant ($\epsilon_e$) of a hypothetic 'liquid of free ions'. Ions in the bound state, on the other hand, are assumed to be composed of cation-anion pairs; this 'ion-pair' assumption (i.e. avoiding special description of larger clusters) dramatically simplifies the formalism.[40] Because of the highly nonuniform charge density of ILs,[7, 41] the dipole moment of an ion in a clustered state is taken to be a fraction of the "full" dipole moment, i.e. $p = m \cdot e \cdot a$, where $a$ is the ion diameter (taken as 1 nm throughout this study), while $m$ a coefficient between 0 and 1 (c.f. Ref. 39). We employ a lattice-gas model to depict such 'ionic liquid' in contact with a charged surface, as illustrated in Figure 1a.

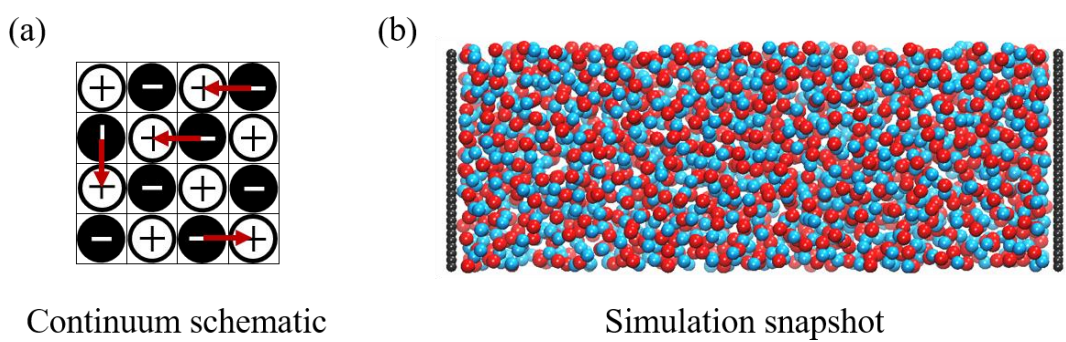

Continuum schematic       Simulation snapshot

Figure 1. (a) Schematics of the lattice-gas description of the ionic liquids consisting of 'free' ions and those bound in ion pairs (red arrow). (b) Snapshot of the MD system with cations (red spheres) and anions (blue spheres). The cation and anion are defined as a bound-state ion pair when their ion centers are within a certain distance from each other, taking such distance to be the sum of the radii of the oppositely charged ions.[36]

The free energy functional for this lattice-gas model can be approximated as[14]

$$F = \int d\mathbf{r} \left( -\frac{1}{2}\epsilon_0\epsilon_e E^2 + e\phi(n_+ - n_-) + F_f(n_+ - n_-) + F_{cl}(n - n_+ - n_-) \right.$$

$$\left. -\frac{1}{2}(n - n_+ - n_-)k_B T \ln \frac{\sinh\left(\frac{pE}{k_B T}\right)}{\frac{pE}{k_B T}} \right. \quad (1)$$

$$\left. - k_B T \ln \frac{n!}{n_+! n_-! \left(\frac{n - n_+ - n_-}{2}\right)!} \right).$$

where $\epsilon_0$ is the vacuum permittivity, $\epsilon_e$ is the dielectric constant of IL constituted of free ions exclusively due to their electronic polarizability, $E$ ($E = -\nabla\phi$) is the electric field, $\phi$ is the electrostatic potential, $F_f$ and $F_{cl}$ are the intrinsic free energy per free ion and clustered ion, respectively, $n_i$ is the number density of free ions (i = +, −) in the unit of nm$^{-3}$ while $n$ is the number density of total lattice sites, $k_B T$ is the thermal energy with $k_B$ and $T$ being the Boltzmann constant and absolute temperature, respectively.

The first two terms in Eq. (1) correspond to the energy of the electrostatic field. From the third and fourth term, we can obtain the fraction of free ions in the bulk IL, $\gamma$, by the following equation analyzed in detail in Ref. 42,

$$F_f - F_{cl} + k_B T \ln\left(\frac{\gamma}{2(1 - \gamma)}\right) = 0. \quad (2)$$

The fifth term accounts for the orientational contribution of ion pairs. The last term in Eq. (1) describes the configurational entropy of the distribution of free ions and ion pairs on the lattice.

The number density of clustered ions, $n_{cl}$, is derived by equalizing the electrochemical potential of each species ($\mu_i = \partial F/\partial n_i$) with their counterparts in the bulk IL (analyzed in detail in the Appendix), and employing the relation $n_{cl} = n - n_+ - n_-$,

$$\frac{n_{cl}}{n} = \frac{(1-\gamma)\left(\frac{\sinh(pE/k_BT)}{pE/k_BT}\right)^{\frac{1}{2}}}{\frac{\gamma}{2}\exp(-e\phi/k_BT) + \frac{\gamma}{2}\exp(e\phi/k_BT) + (1-\gamma)\left(\frac{\sinh(pE/k_BT)}{pE/k_BT}\right)^{\frac{1}{2}}}. \quad (3)$$

A detailed derivation of $n_+$ and $n_-$, following the approach of Ref. 19, is shown in the Appendix.

The modified Poisson equation and the expression for dielectric constant are obtained by substituting the free energy functional into Euler-Lagrange equation $\frac{\partial}{\partial x}\frac{\partial f}{\partial \phi'} - \frac{\partial f}{\partial \phi} = 0$, analyzed in detail in the Appendix, and are given by,

$$\frac{d}{dx}\left(\epsilon_0 \epsilon_d \frac{d\phi}{dx}\right) = -e(n_+ - n_-), \quad (4)$$

$$\epsilon_d = \epsilon_e + \frac{(n - n_+ - n_-)pL(pE/k_BT)}{2\epsilon_0 E}. \quad (5)$$

where $L(s) = (\coth(s) - 1/s)$ is the Langevin function.[31] The first term in Eq. (5) describes the electronic degrees of freedom of ions, and the second term originates from the orientational ordering of ion pairs, and through the factor $(n - n_+ - n_-)$, it is affected by the interchange between free and bound states in EDL.

*Molecular dynamics simulations*

As shown in Figure 1b, our MD simulation system consists of two identical electrodes with a slab of IL enclosed between them. The distance between two electrodes is 30 nm, which is sufficiently large to ensure electroneutrality and bulk-like IL behavior in the middle of the system (not perturbed by the electrodes). The force fields of the electrodes and ILs are taken from Ref. 43; each electrode is made of Lennard-Jones (LJ) spheres arranged in a square lattice with a lattice spacing of 0.33 nm. The cations and anions of ILs are modeled as symmetrical LJ spheres with 1 nm diameter and opposite unit charge, i.e. the model of ions is also made as simple as possible to be able to compare it with the Coulomb lattice gas theory. The cation and anion are defined as a bound-state ion pair when their ion centers are within a certain distance from each other, and such distance was taken to be the sum of the radii of the oppositely charged ions.[36]

Simulations were performed in the NVT ensemble using the GROMACS package.[44] The temperature was maintained at 450K with Nosé-Hoover thermostat (the temperatures where elevated because ideal, identical size charged Lennard-Jones spheres tend to freeze at room temperature). Periodic boundary conditions in all three directions were used. In order to eliminate artifacts of the periodicity in the direction perpendicular to the electrodes, the length of the simulation box in this direction was set to be three times the width between the electrodes. The equilibration was performed for 5 ns with time step of 0.01 ps, following by another 20 ns production for further analysis.

The electrical potential distribution was calculated as,

$$\phi(z) = -\frac{\sigma}{\epsilon_0 \epsilon_e} z - \frac{1}{\epsilon_0 \epsilon_e} \int_0^z (z - z') \rho(z') \, dz'. \tag{6}$$

Here $\sigma$ is the surface charge density, $\epsilon_e$ is assigned value 2 in our simulations, and $\rho$ is the ionic charge density along the direction perpendicular to the electrodes. Thus, the potential drop across the EDL ($\phi_{EDL}$) is calculated relative to the potential of zero charge (PZC).[45]

$$\phi_{EDL} = (\phi_{electrode} - \phi_{bulk}) - (\phi_{electrode} - \phi_{bulk})|_{PZC}. \tag{7}$$

where $\phi_{electrode}$ and $\phi_{bulk}$ is, respectively, the potential on the electrode surface and in the bulk.

## Results and discussion

*Number density profiles*

In what follows, we articulate the theoretical prediction in terms of the number density of clustered ions, $n_{cl}$, and then compare it with our MD simulations. We set $n_{cl}/n$ in bulk IL from theory the same as that from simulations. In this case, $n_{cl}/n$ in bulk IL are both 0.54. The dipole moment of an ion pair is set as $8.8 \times 10^{-29}$ C m ($0.45 \times ea$) in order that $\epsilon_d$ at bulk IL is around 15.[7] Figure 2a,b show $n_{cl}$ at different electrode potentials. Because of the simplified structure and monovalence of the modeled IL in both theory and simulations, the applied potential range must not be confined within the typical electrochemical window of ILs,[46, 47] but rather broader to manifest some interesting characteristics, so the reader should not be worried by large voltages that we will handle in this academic study. Firstly, we focus on the $n_{cl}$ profile at -8 V from the theory side, where it first increases and then decreases from the bulk to the electrode. The hump in the $n_{\text{cl}}$ profile is attributed to the presence of two competing forces affecting clustered ions: they tend to exist in regions of higher electric fields because of polarization forces, however, sufficiently high electric fields ultimately unbind the clustered ions into monopoles.

In addition, Figure 2a shows that as the electrode accumulates a higher charge, the unbinding of ion pairs causes the hump of $n_{\text{cl}}$ profile to reside further away from the electrode, which, as depicted in Figure 2b, is corroborated by MD simulations. Particularly, the locations of the humps at -0.5 V (blue curve) and -5.5 V (orange curve) obtained from theory are in good agreement with those from MD simulations. It is worth reiterating that because of the mean-field nature of our theory, the layered structure inherently revealed by the MD simulation cannot be captured by our theory, which explains why the single hump in theory continuously moves toward the bulk as the electrode potential gets higher biased, whereas in simulations we see a layered structure of several humps, and their location is weakly affected by the electrode potential.[48]

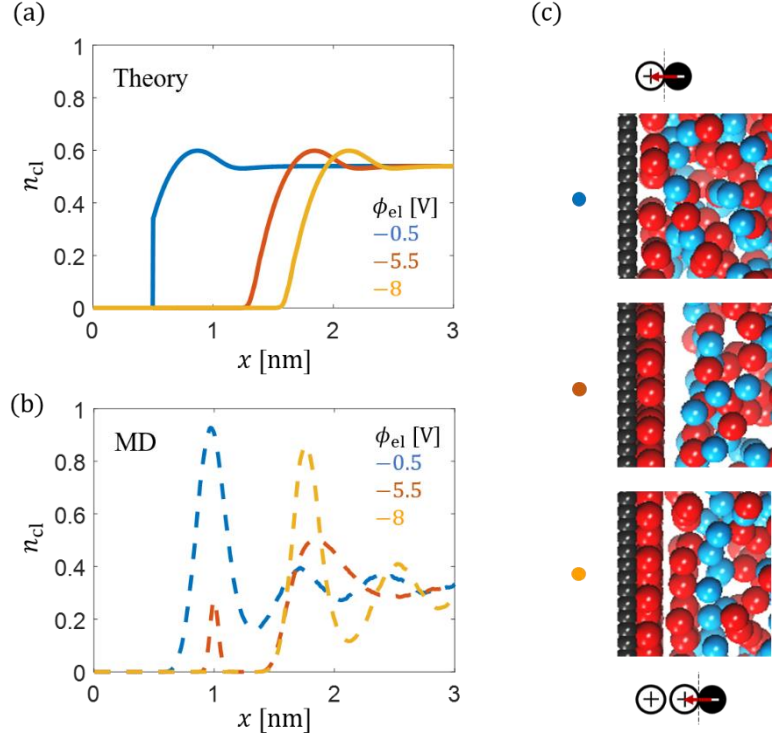

Figure 2. (a-b) The number density profiles of ion pairs at -0.5 V, -5.5 V and -8 V obtained from theory (a) and MD (b), respectively. $T =$ 450 K, $n_{cl}/n$ of bulk IL is 0.54. For the theory side, $p = 0.55 \times ea = 8.8 \times 10^{-29}$ C m. For the MD side, the total number density of ions, both free and clustered, in bulk IL is 0.62 nm$^{-3}$ (lower than 1 nm$^{-3}$ from the theory side). The fraction of clustered ions in bulk IL is 0.54 and, therefore the corresponding $n_{cl}$ is 0.34 nm$^{-3}$. (c) Snapshots of the MD simulations performed at -0.5 V (blue dot), -5.5 V (orange dot) and -8 V (yellow dot).

Figure 3 displays the theoretically-obtained effective dielectric constant [$\epsilon_d$, Eq. (5)] profile at -0.5, -5.5 and -8 V. It is shown that $\epsilon_d$ experiences a drastic decrease from 15 to almost 2 as approaching the electrode. This stems from the prediction of the model that clustered ions unbind in the EDL, and therefore the dielectric screening is reduced.

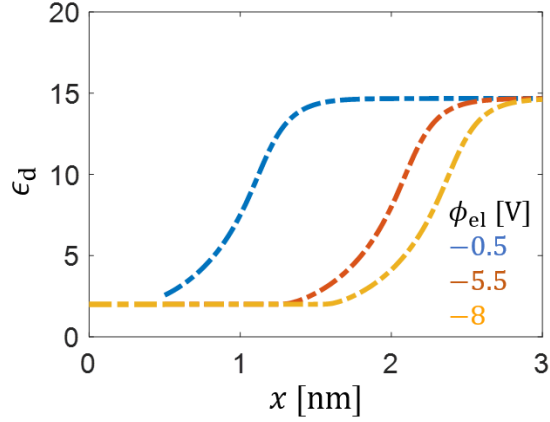

Figure 3. Dielectric constant profiles in the vicinity of the electrode with different potentials obtained from theory. The parameters are set to be the same as those in Figure 2.

*Differential capacitance*

The traditional wisdom in the IL-based EDL theory articulates that the differential capacitance curve in the primitive model displays a camel shape if the fraction of free charge carriers in the bulk electrolyte is lower than $1/3$.[14] However, the incorporation of the clustered ions in the present model changes the threshold ($\gamma = 1/3$) of the "camel-bell transition". In this section, we investigate the $C$-$\phi_{el}$ curve of the case where $\gamma$ is 0.2 (lower than $1/3$), and see if it displays a camel shape, as predicted by the primitive model. For both models, we set $\epsilon_d$ as 10.5 in bulk IL and therefore the dipole moment of the present model is $4.8 \times 10^{-29}$ C m ($0.3 \times ea$). As revealed by Figure 4a, the present model may well show a bell-shaped $C - V$ curve despite $\gamma$ is lower than the $1/3$ threshold, contrary to the camel-shaped dashed curve predicted by the primitive model employed in Ref. 14. The model, of course, could give a $C$-$\phi_{el}$ curve of camel shape, but the corresponding number density would be lower than $1/3$ and dependent of the magnitude of $p$. On the other hand, the model gives a lower $C$ than the primitive model,[14] which is explained by the reduction in the dielectric screening depicted in Figure 4b. No matter how the electrode is charged, $\epsilon_d$ remains unaltered in the primitive model. In the present model, however, clustered ions near the electrode-IL interface shift to free state, resulting a decreased $\epsilon_d^{\text{surf}}$.

This finding serves as a reminder for us when interpreting experimental data: a bell-shaped $C - V$ curve from experiment is not a certain indicator of highly dissociated electrolyte, instead, it may be a result of the formation and break-down of ion clustering.

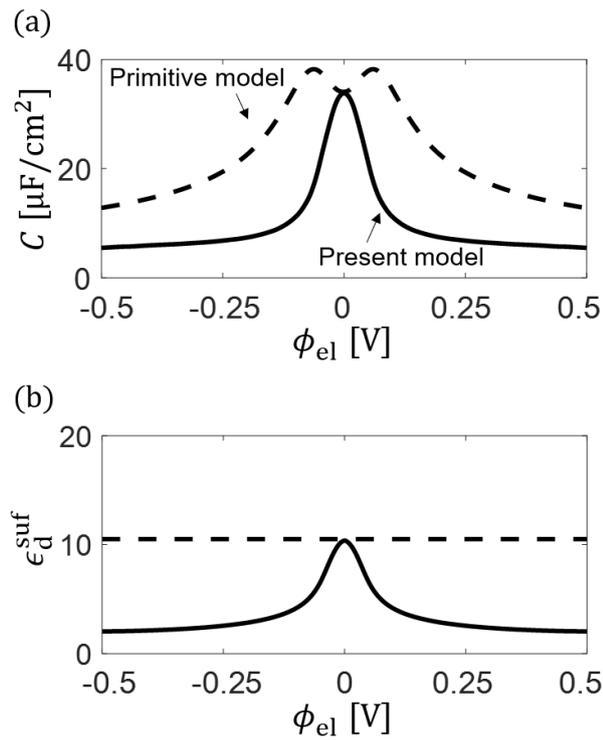

Figure 4. (a) Differential capacitance curves of the primitive model (dashed line) and present model (solid line).[14] In both cases, $\gamma = 0.2$. In the present model, $p = 0.3 \cdot ea = 4.8 \cdot 10^{-29}$ C m. $T = 300$ K. (b) Dielectric constant vs electrode potential curve. Dashed line: primitive model; solid line: present model. In both cases, $\gamma = 0.2$. For both models, $\epsilon_d$ is 10.5 in the bulk.

## Conclusions

In this work, we proposed a mean-field lattice-gas theory of the EDL in an IL-system with transient clustered ions bearing dipole moments. Using the analytical expression for the spatially-varying dielectric constant, we obtained the number density distribution of clustered ions which was compared with our MD simulations. They both showed a shift of ions from clustered state to free state in high electric field. The resulting decrement of dielectric constant at the interface led to the reduction of the threshold of "camel-bell transition"; that is, even if the proportion of free charge carrier is lower than the 'canonical threshold', 1/3, the differential capacitance vs potential curve may well display a bell shape.

As mentioned in the introduction, the model explored here is crude. Even using the concept of local dielectric constant at such distances is problematic (see multiple works on the nonlocal electrostatic theory of the electrical double layer).[49, 50] But as a qualitative signature of the effect, it sounds convincing: ions are liberated into free states by the electric field inside the EDL which crashes the ion clusters, for mere simplicity considered here as cation-anion pairs.

## Acknowledgement

Y.Z. expresses his sincere gratitude to Dechun Si, Honghui Lv, Pedro de Souza and Xiaoyue Wang for discussions, suggestions, inspirations and especially encouragement. T.Y., M.C., and G.F. acknowledge the funding support from the National Natural Science Foundation of China (51876072). A.A.K. and Z.A.H.G. acknowledges discussion of various related issues with Michael Patrick McEldrew and Martin Bazant. J.H. acknowledges financial support from National Natural Science Foundation of China (21802170). Participation of Z.A.H.G in this work was supported through a studentship of the Centre for Doctoral Training on Theory and Simulation of Materials at Imperial College London, funded by the EPSRC (EP/L015579/1), and the funding from the Thomas Young Centre under grant number TYC-101. All MD simulations were performed at the National Supercomputing Centers in Guangzhou (Tianhe II).

## Conflict of Interest

The authors declare no conflict of interest.

## Appendix

Taking the variation of $f$ with respect to $n_i$ yields dimensionless electrochemical potential of cations and anions:

$$\mu_+ = \frac{e\phi}{k_\mathrm{B}T} + \frac{1}{2}\ln\frac{\sinh\left(\frac{pE}{k_\mathrm{B}T}\right)}{\frac{pE}{k_\mathrm{B}T}} - \ln\frac{\left(\frac{n - n_+ - n_-}{2}\right)^{\frac{1}{2}}}{n_+}, \quad (8)$$

$$\mu_- = -\frac{e\phi}{k_\mathrm{B}T} + \frac{1}{2}\ln\frac{\sinh\left(\frac{pE}{k_\mathrm{B}T}\right)}{\frac{pE}{k_\mathrm{B}T}} - \ln\frac{\left(\frac{n - n_+ - n_-}{2}\right)^{\frac{1}{2}}}{n_-}. \quad (9)$$

In the bulk, the electric field is totally screened and the electrostatic potential is taken to be zero; and we have $n_+ = n_- = n_0 = \gamma n/2$ due to electric neutrality where $n_0$ denotes the number density of cations (or anions) in the bulk.

Equalizing the electrochemical potential of each species to its counterpart in the bulk electrolyte, we obtain

$$\frac{e\phi}{k_BT} + \frac{1}{2}\ln\frac{\sinh\left(\frac{pE}{k_BT}\right)}{\frac{pE}{k_BT}} - \ln\frac{n_0}{n_+}\left(\frac{n-n_+-n_-}{n-n_0-n_0}\right)^{\frac{1}{2}} = 0, \qquad (10)$$

$$-\frac{e\phi}{k_BT} + \frac{1}{2}\ln\frac{\sinh\left(\frac{pE}{k_BT}\right)}{\frac{pE}{k_BT}} - \ln\frac{n_0}{n_-}\left(\frac{n-n_+-n_-}{n-n_0-n_0}\right)^{\frac{1}{2}} = 0. \qquad (11)$$

Trivial rearrangements give

$$\exp\left(\frac{e\phi}{k_BT}\right)\left(\frac{\sinh\left(\frac{pE}{k_BT}\right)}{\frac{pE}{k_BT}}\right)^{\frac{1}{2}} = \frac{n_0}{n_+}\left(\frac{n-n_+-n_-}{n-n_0-n_0}\right)^{\frac{1}{2}}, \qquad (12)$$

$$\exp\left(-\frac{e\phi}{k_BT}\right)\left(\frac{\sinh\left(\frac{pE}{k_BT}\right)}{\frac{pE}{k_BT}}\right)^{\frac{1}{2}} = \frac{n_0}{n_-}\left(\frac{n-n_+-n_-}{n-n_0-n_0}\right)^{\frac{1}{2}}. \qquad (13)$$

To simplify further derivations, we will use a kind of interpolation approximation to the right-hand sides of these two equations. We brutally omit the square root there. Indeed the $\left(\frac{n-n_+-n_-}{n-n_0-n_0}\right)^{\frac{1}{2}}$ and $\frac{n-n_+-n_-}{n-n_0-n_0}$ have the same limiting behaviors: both are 0 when $n_+ + n_- = n$, and both are equal to 1 when $n_+ + n_- = 2n_0$. In between, this function would of course, be different, but the difference is not large and will not qualitatively affect the results. Such an approach has been used, in Ref. 19. We can then get simple analytical expressions for $n_+$ and $n_-$

$$\frac{n_+}{n} = \frac{n_0 \exp\left(-\frac{e\phi}{k_B T}\right)}{n_0 \exp\left(-\frac{e\phi}{k_B T}\right) + n_0 \exp\left(\frac{e\phi}{k_B T}\right) + (n - n_0 - n_0)\left(\frac{\sinh\left(\frac{pE}{k_B T}\right)}{\frac{pE}{k_B T}}\right)^{\frac{1}{2}}}, \quad (14)$$

$$\frac{n_-}{n} = \frac{n_0 \exp\left(\frac{e\phi}{k_B T}\right)}{n_0 \exp\left(-\frac{e\phi}{k_B T}\right) + n_0 \exp\left(\frac{e\phi}{k_B T}\right) + (n - n_0 - n_0)\left(\frac{\sinh\left(\frac{pE}{k_B T}\right)}{\frac{pE}{k_B T}}\right)^{\frac{1}{2}}}. \quad (15)$$

The number density expression for ion pairs is given by $n_{cl} = n - n_+ - n_-$ and is shown by Eq. (3).

*Derivation of the modified Poisson equation and dielectric constant*

Now, we develop the modified Poisson equation as well as the expression for dielectric constant. Substituting the free energy functional into Euler-Lagrange equation $\frac{\partial}{\partial x}\frac{\partial f}{\partial \phi'} - \frac{\partial f}{\partial \phi} = 0$, we obtain

$$\epsilon_0 \epsilon_e \frac{dE}{dx} + \frac{d}{dx}\left(\frac{1}{2}(n - n_+ - n_-)pL(pE/k_B T)\right) = e(n_+ - n_-), \quad (16)$$

where $L(s)$ is the Langevin function, defined in the main text. Rearrangement gives

$$\frac{d}{dx}\left(\epsilon_0\left(\epsilon_e + \frac{(n - n_+ - n_-)pL\left(\frac{pE}{k_B T}\right)}{2\epsilon_0 E}\right)E\right) = e(n_+ - n_-), \quad (17)$$

Finally, we obtain the modified Poisson equation

$$\frac{d}{dx}\left(\epsilon_0\epsilon_d \frac{d\phi}{dx}\right) = -e(n_+ - n_-). \tag{18}$$

where $\epsilon_d$ denotes the electric-field-dependent dielectric constant of the system, given by

$$\epsilon_d = \epsilon_e + \frac{(n - n_+ - n_-)pL\left(\frac{pE}{k_BT}\right)}{2\epsilon_0 E}. \tag{19}$$